\documentclass[aps,showpacs,amsmath,prepintnumbers,amssymb,superscriptaddress,reprint]{revtex4-2}

\usepackage{epsf}
\usepackage{graphicx}
\usepackage{color}
\usepackage[normalem]{ulem}
\usepackage{newfloat}
\usepackage[latin1]{inputenc}
\usepackage{comment}
\usepackage{amsmath}
\usepackage{physics}
\usepackage[dvipdfmx]{hyperref}
\usepackage{float}
\usepackage[version=4]{mhchem}

\DeclareFloatingEnvironment[name={Supplementary-Figure}]{suppfigure}

\begin{document}
\title{Zone-selection effect of photoelectron intensity distributions \\in a nonsymmorphic system \textbf{\textit{R}}AlSi (\textbf{\textit{R}} : Ce and Nd)}

\author{Yusei Morita}
\affiliation{Department of Physics, Faculty of Science, Hiroshima University, Higashi-hiroshima, Hiroshima 739-8526, Japan}

\author{K. Nakanishi} 
\affiliation{Graduate School of Advanced Science and Engineering, Hiroshima University, Higashi-hiroshima, Hiroshima 739-8526, Japan}

\author{T. Iwata} 
\affiliation{Graduate School of Advanced Science and Engineering, Hiroshima University, Higashi-hiroshima, Hiroshima 739-8526, Japan}
\affiliation{International Institute for Sustainability with Knotted Chiral Meta Matter (WPI-SKCM$^2$), Hiroshima University, Higashi-hiroshima, Hiroshima 739-8526, Japan}

\author{K. Ohwada} 
\affiliation{Graduate School of Advanced Science and Engineering, Hiroshima University, Higashi-hiroshima, Hiroshima 739-8526, Japan}

\author{Y. Nishioka}
\affiliation{Graduate School of Advanced Science and Engineering, Hiroshima University, Higashi-hiroshima, Hiroshima 739-8526, Japan}

\author{T. Kousa}
\affiliation{Graduate School of Advanced Science and Engineering, Hiroshima University, Higashi-hiroshima, Hiroshima 739-8526, Japan}

\author{M. Nurmamat} 
\affiliation{Graduate School of Advanced Science and Engineering, Hiroshima University, Higashi-hiroshima, Hiroshima 739-8526, Japan}

\author{K. Yamagami} 
\affiliation{Synchrotron Radiation Research Center, Japan Atomic Energy Agency, Hyogo 679-5148, Japan}

\author{A. Kimura} 
\affiliation{Graduate School of Advanced Science and Engineering, Hiroshima University, Higashi-hiroshima, Hiroshima 739-8526, Japan}
\affiliation{International Institute for Sustainability with Knotted Chiral Meta Matter (WPI-SKCM$^2$), Hiroshima University, Higashi-hiroshima, Hiroshima 739-8526, Japan}
\affiliation{Research Institute for Semiconductor Engineering, 1-4-2 Kagamiyama, Higashi-Hiroshima, Hiroshima 739-8527, Japan}

\author{T. Yamada} 
\affiliation{Liberal Arts and Sciences, Faculty of Engineering, Toyama Prefectural University, Izumi, Toyama 939-0398, Japan}

\author{H. Tanida} 
\affiliation{Liberal Arts and Sciences, Faculty of Engineering, Toyama Prefectural University, Izumi, Toyama 939-0398, Japan}

\author{Kenta Kuroda} 
\email{kuroken224@hiroshima-u.ac.jp}
\affiliation{Graduate School of Advanced Science and Engineering, Hiroshima University, Higashi-hiroshima, Hiroshima 739-8526, Japan}
\affiliation{International Institute for Sustainability with Knotted Chiral Meta Matter (WPI-SKCM$^2$), Hiroshima University, Higashi-hiroshima, Hiroshima 739-8526, Japan}
\affiliation{Research Institute for Semiconductor Engineering, 1-4-2 Kagamiyama, Higashi-Hiroshima, Hiroshima 739-8527, Japan}

\date{\today}
\begin{abstract}
We investigate the electronic structures of noncentrosymmetric Weyl semimetals $R\rm{AlSi}$ ($R$: Ce and Nd) using soft x-ray angle-resolved photoemission spectroscopy.
We find that the photoelectron intensity distribution observed in the momentum-resolved electronic bands is highly sensitive to the covered Brillouin zone (BZ) due to the zone-selection effect arising from the nonsymmorphic crystal structure of $R\rm{AlSi}$.
Our data reconstruct the photoelectron distributions varied according to the zone-selection effect, and reveal comprehensive information about the electronic band structures reproduced by band calculations.
This detailed information enables us to experimentally trace the Weyl-cone dispersion throughout three-dimensional momentum space, providing valuable insights into the unique properties of $R\rm{AlSi}$.
\end{abstract}
\maketitle

In the last decade, a great richness of quantum states has been uncovered in topological semimetals~\cite{Weng_jpcm2016,binghai_arcmp2017, armitage_rmp2018, Ding_rmp2021}.
Among them, Dirac semimetal is known as the most representative example, where Dirac fermions manifest as low-energy quasiparticle excitations near the crossing point of the linear band dispersion~\cite{Borisenko_prl14, Xu_Science15_Dirac, Liu_NatureMat14}.
Starting from this prototypical case, breaking of either space-inversion or time-reversal symmetry results in the lifting off the degeneracy of the Dirac points, giving rise to the emergence of pairs of Weyl points~\cite{binghai_arcmp2017, armitage_rmp2018}.
Weyl semimetals in the non-magnetic context have been intensively studied in noncentrosymmetric materials~\cite{Xu_Science15,Lv_prx15} by angle-resolved photoemission spectroscopy (ARPES)~\cite{Damascelli_rmp03,Sobota_rmp2021,Zhang_NatureRev22}.
In particular, the magnetic counterpart of Weyl semimetals is attractive, because it is adequate to control the distribution of Weyl points and large fictitious fields by selecting magnetic textures in real space, which shows their application potential for the next-generation spintronics~\cite{Nakatsuji_ANN2022}.
However, the experimental identifications of magnetic Weyl semimetals remains challenging, primarily due to difficulties in observing low-energy excitation states in magnetic materials~\cite{Kuroda_NM2017, Liu_SciAdv19}.

A family of $R\rm{Al}X$ ($R$=rare earth elements, $X$=Si and Ge) compounds represents a promising candidate for magnetic Weyl semimetals.
These materials exhibit a noncentrosymmetric structure with space group $I4_{1}md$~\cite{Hodovanets_prb18, Su_prb21, yang2021noncollinear}, categorized as a nonsymmorphic space group.
In this structure, the rare-earth ions form two interpenetrating body-centered-tetragonal sublattices offset by (0, $a$/2, $c$/4) [as depicted in Fig.~\ref{fig1}(a)].
Due to the breaking of inversion symmetry, the series material can exhibit Weyl semimetal phases in the paramagnetic phase~\cite{sciadv.1603266, Sanchez:2020aa, PhysRevB.107.035158, PhysRevMaterials.7.L051202, Zhang:2023aa, Li:2023aa, PhysRevB.109.035120, Cheng:2024aa}.
Among them, CeAlSi (NdAlSi) exhibits ferromagnetic (antiferromagnetic) ordering below $T_{\rm{C}}$ ($T_{\rm{N}}$) =~8.5~K (7.2~K), leading to anomalous angular magnetoresistance~\cite{suzuki2019singular}, spin density waves~\cite{gaudet2021weyl}, chiral domain-walls~\cite{sun2021mapping} and anisotropic anomalous Hall effect~\cite{yang2021noncollinear}.
These properties are believed to originate from Weyl fermions in the magnetic order.
However, the electronic structure of $R\rm{Al}X$ compounds has not been fully elucidated yet, primarily because Weyl nodes are present three-dimensionally in momentum space~\cite{Guoqing_prb18}.
According to first-principles calculations, there are three types of thirty-two Weyl nodes (eight W1, sixteen W2, and eight W3 nodes) in the vicinity of the Fermi energy ($E_{\rm{F}}$) in the Brillouin zone, as depicted in Fig.~\ref{fig1}(b).
To fully characterize all types of Weyl nodes, comprehensive ARPES measurements covering three-dimensional momentum space are required.

In this Letter, we employ angle-resolved photoemission spectroscopy combined with synchrotron soft x-ray (SX-ARPES) to comprehensively investigate the electronic structures of CeAlSi and NdAlSi. 
We observe strong zone selection effect that significantly influences the photoelectron distributions, thereby complicating the interpretation of intrinsic band structures.
By covering a number of the Brillouin zones (BZs) information, we fully reconstruct the photoelectron distribution.
The reconstructed distribution displays three-dimensional electronic structures, all of which are well reproduced by our density functional theory (DFT) calculations.
Eliminating the zone selection effect provides us with a comprehensive understanding of the electronic properties and enables us to trace the Weyl-cone dispersions throughout the three-dimensional momentum space.

Single crystalline samples of $R\rm{AlSi}$ were synthesized by an Al-self flux method using an alumina crucible, sealed in a silica tube with an Ar-gas atmosphere.
SX-ARPES measurement was performed at BL25SU in SPring-8~\cite{Muro_jsr2021}.
Photoelectrons were acquired by the hemispherical analyzer ScientaOmicron DA30.
The total experimental energy resolution was set to approximately 80 meV for photon energies ($h\nu$) ranging from 560 to 800~eV.
Clean surfaces were obtained by cleaving in a ultra-high vacuum with a pressure of $3.5\times 10^{-8}$ Pa.
The sample temperature was kept around 60~K during measurements well above the magnetic transition temperatures.
Band calculations were performed using the DFT-based $ab$ $initio$ calculation package WIEN2k~\cite{Blaha2020}, employing the PBE-GGA exchange-correlation potential and spin-orbit coupling.
We used the experimental lattice parameters, with lattice constants $a=b=4.25~{\rm \AA}$ and $c=14.5~{\rm \AA}$.
WannierTools~\cite{Wu2017} has been used to identify the Weyl points in the Brillouin zone.

Let us start by presenting ARPES results of CeAlSi obtained with various SX photon energies ($h\nu{'}s$).
Figure~\ref{fig1}(c1) presents the $k_{z}$-$k_{x}$ constant energy contours at $k_y$=0~${\rm{\AA}}^{-1}$ at $E-E_{\rm{F}}$=$-$1.18~eV recorded with varying $h\nu{'}$s from 560 to 780~eV.
This dataset covers multiple BZs including a number of high-symmetry momentum points: $\Gamma$, $\rm{Z}$ and $\Sigma$ points.
We observe elliptical contours surrounding $\Sigma$ points, which unambiguously indicate that the observed band features belong to bulk bands rather than surface states that do not disperse along the $k_z$ direction.
The signature of $k_z$ dispersion is more pronounced in Figs.~\ref{fig1}(d1) and \ref{fig1}(d2) where we show the $E$-$k_z$ band map at $k_x$ =$-$0.33~${\rm{\AA}}^{-1}$ [the dashed line in Fig.~\ref{fig1}(c1)] and the representative energy distribution curves (EDCs). 
From these data, it is evident that the observed bands disperse along the $k_z$ direction and exhibits the band folding with a periodicity of ${\Delta}k_z\sim$0.86~${\rm{\AA}}^{-1}$  [guided by arrows in Fig.~\ref{fig1}(d2)] that is consistent with 2${\pi}/c$ ($c$ is the lattice constant of the primitive cell).
From these signatures, we conclude that our data detect the bulk electronic structures thanks to SX photon sources (see also Supplementary~Note~1 for NdAlSi).

Interestingly, we observe not only $k_z$-dispersion but also periodic modulation of photoelectron intensity in momentum space.
One notable signature is the considerable difference in the shape of the momentum distribution curves (MDCs) at two $\Gamma$ points ($\Gamma_1$ and $\Gamma_2$) despite these two $k_z$-levels being equivalent in BZ [Figs.~\ref{fig1}(c2) and \ref{fig1}(c3)].
In the MDC at the $\Gamma_2$-level [the blue curve in Fig.~\ref{fig1}(c3)], the photoelectron intensity is predominantly distributed around $k_x = \pm$0.7~${\rm{\AA}}^{-1}$, whereas at the $\Gamma_1$-level, it is distributed around $k_x = \pm$0.2~${\rm{\AA}}^{-1}$ [the red curve in Fig.~\ref{fig1}(c3)].
In general, photoelectron intensity is sensitive to excitation $h\nu$ due to the matrix element of optical transitions as well as the photoemission scattering process, known as final state effect~\cite{Damascelli_rmp03}.
However, our results cannot be explained solely by the conventional $h\nu$ dependence.
In stark contrast to the results at the $\Gamma_1$- and $\Gamma_2$-levels, the MDCs at each $\rm{Z}$ point exhibit similar shapes even though these are acquired by different $h\nu{'}$s.
Therefore, the distribution of photoelectron intensity appears to be more influenced by the nature of BZ rather than by the excitation $h\nu$ (namely final state effect).

\begin{figure}[t!]
\begin{center}
\includegraphics[width=1\columnwidth]{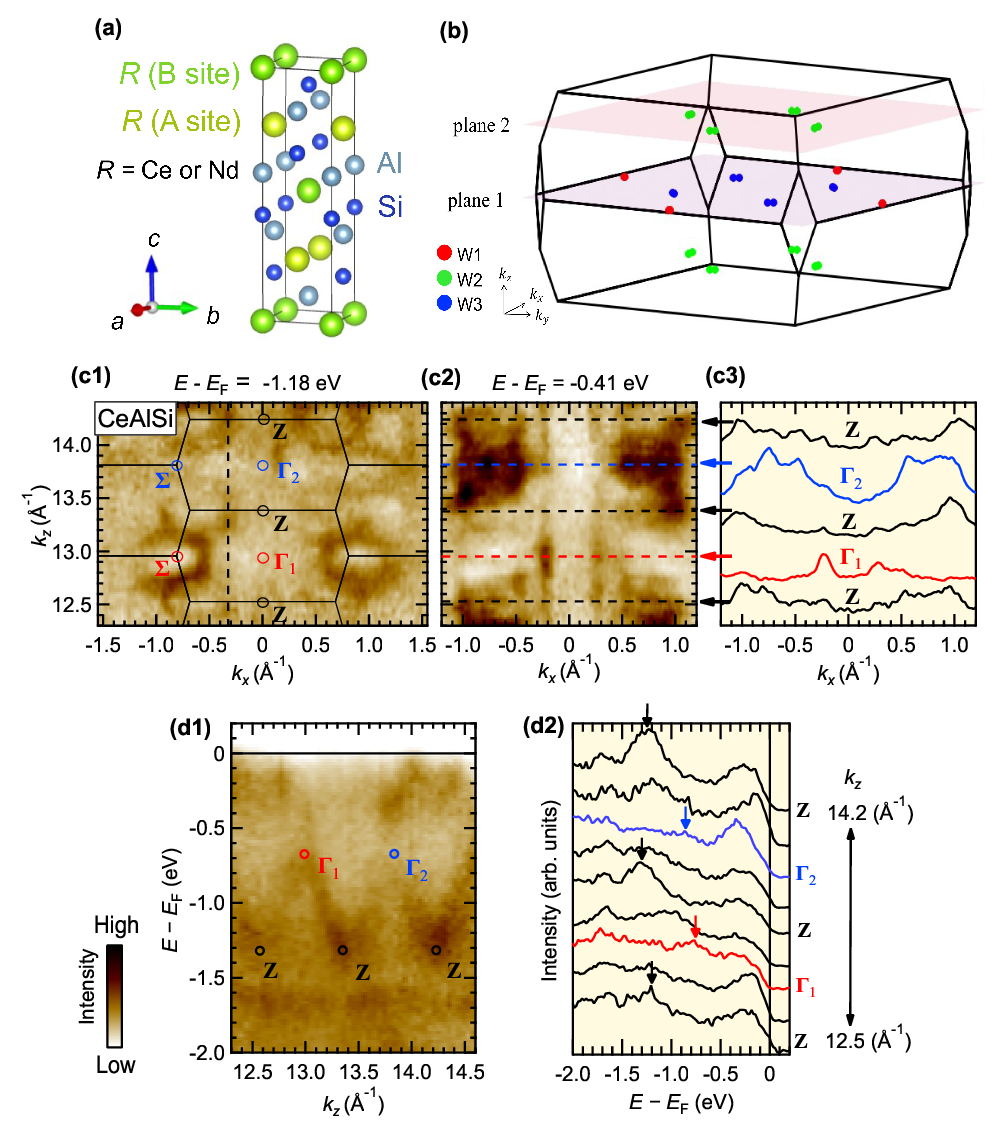}
\caption{
(a) Crystal structure of $R$AlSi ($R$: Ce and Nd).
(b) Distribution of the Weyl nodes in three-dimensional BZ predicted by DFT calculations.
The two $k_x$-$k_y$ planes (plane~1 and plane~2) for their observations by our SX-ARPES (see Fig.~\ref{fig4}).
The $k_z$ values of plane~1 and plane~2 correspond to $k_z$=0 and 0.3~$\rm{\AA}^{-1}$, respectively.
(c1) and (c2) $k_z$-$k_x$ constant energy contour mapping of CeAlSi at $E-E_{\rm{F}}$=$-$1.18~eV and $-$0.41~eV, scanned by varying $h\nu{'}$s from 560 to 780~eV with 5~eV step.
The inner potential $V_0$=17~eV is used to convert $h\nu$ dependency of the experimental bands to corresponding $k_z$ dispersion.
The boundary of the BZs are drawn by solid lines.
(c3) The momentum distribution curves of (c2) at representative $k_z$-levels. 
(d1) The band map cut on the $E$-$k_z$ plane at $k_{x}$=$-$0.43$~\rm{\AA^{-1}}$ [the dashed line in (c1)].
(d2) The representative energy distribution curves.
The periodic $k_z$-dispersion is traced by the arrows.  
}
\label{fig1}
\end{center}
\end{figure}

\begin{figure*}[t!]
\begin{center}
\includegraphics[width=1.0\textwidth]{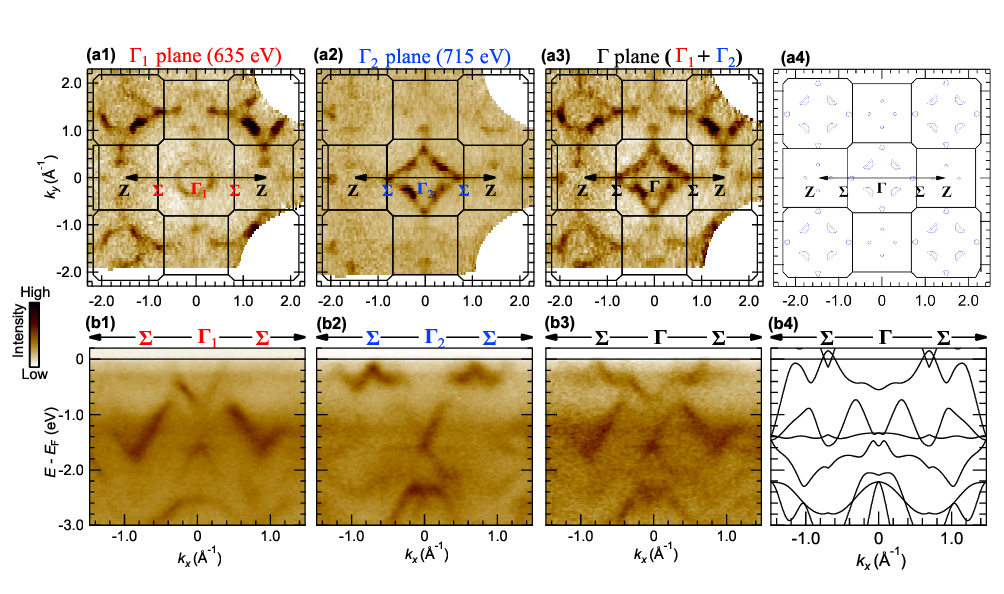}
\caption{
(a1), (a2) The results of the Fermi surface mapping in NdAlSi on the $k_x$-$k_y$ planes cutting $\Gamma_1$ ($h\nu$=635~eV) and $\Gamma_2$ ($h\nu$=720~eV).  
(a3) The reconstructed photoelectron intensity distribution obtained by summing the intensities of the FS maps [(a1) and (a2)], where the data from (a1) and (a2) were weighted with ratios of 1:0.8.
(a4) The calculated Fermi surface at $\Gamma$.
(b1)-(b3) The photoelectron distributions cut along $\Sigma$-$\Gamma$-$\Sigma$ line for data shown in (a1)-(a3), compared to the calculated band dispersions shown in (b4).
We shift the overall DFT bands in energy of 80 meV to obtain the best fit.
}
\label{fig2}
\end{center}
\end{figure*}

The periodic photoelectron intensity modulation is observed also in $k_x$-$k_y$ Fermi surface (FS) contours cutting either $\Gamma_1$ or $\Gamma_2$ point as presented in Figs.~\ref{fig2}(a1) and \ref{fig2}(a2).
A diamond-shaped intensity distribution is clearly discernible on the $\Gamma_2$-plane surrounding normal emission ($k_x$=$k_y$=0~$\rm{\AA}^{-1}$) [Fig.~\ref{fig2}(a2)].
However, at away from normal emission, the intensity diminishes at the other covered $\Gamma$ points.
Conversely, the intensity around normal emission on the $\Gamma_1$-plane is relatively weak, while it becomes stronger at the other covered $\Gamma$ points [Fig.~\ref{fig2}(a1)].
Since the electronic band dispersion itself should reflect the periodicity of BZ and hence should be equivalent at all $\Gamma$ points, the observed periodic modulation of photoelectron intensity distribution cannot be explained by the band structure alone.

\begin{figure}[t!]
\begin{center}
\includegraphics[width=1\columnwidth]{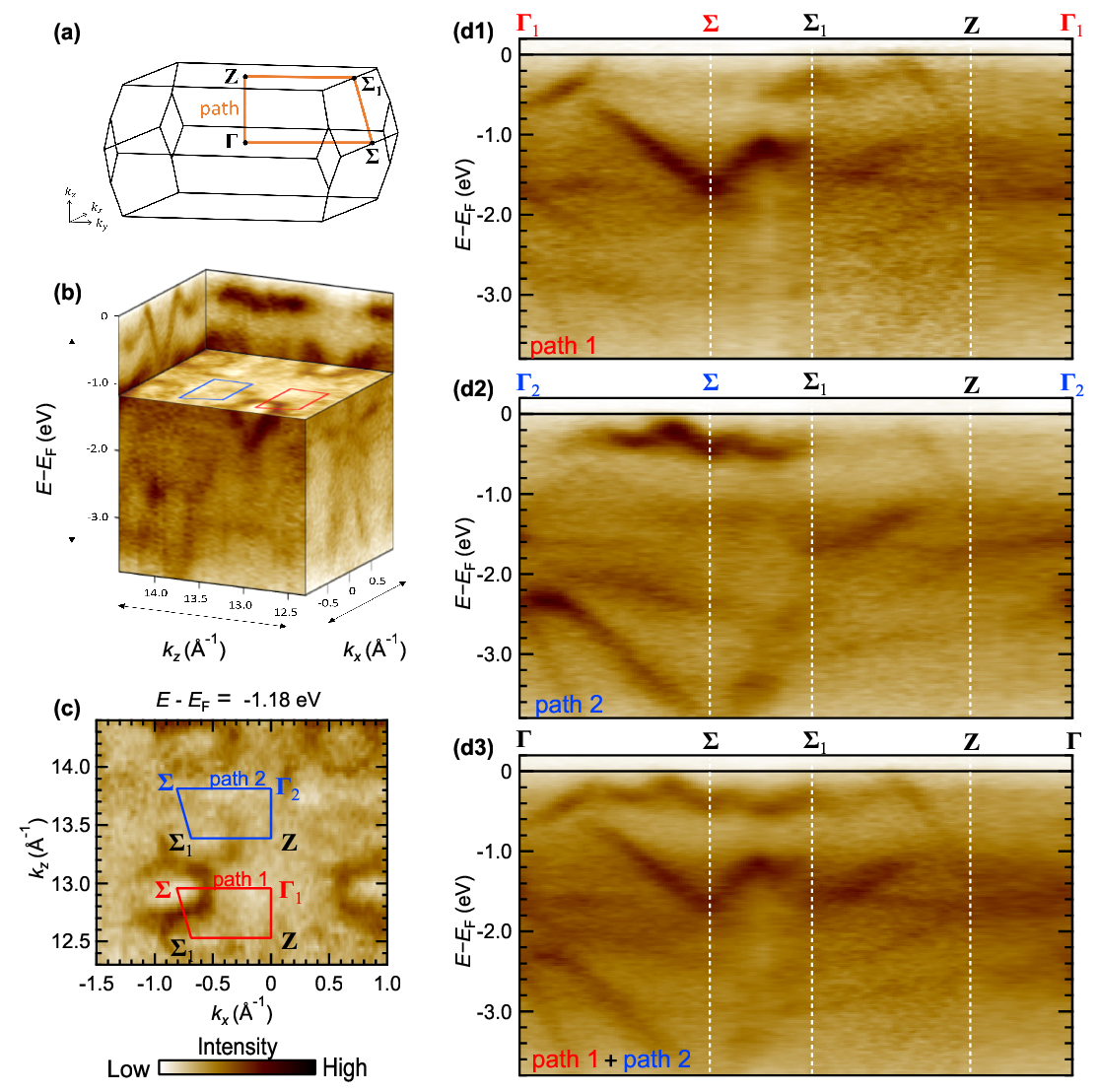}
\caption{
(a) High-symmetry $k$-path in a BZ.
(b) Multi-dimensional ARPES data including the photoelectron intensity as a function of $E$, $k_z$ and $k_x$.
(c) The $k_z$-$k_x$ map at $E-E_{\rm{F}}$=$-1.18$~eV includes the two equivalent $k$-paths corresponding to the paths in (a).
(d1) and (d2) $E$-$k$ map cut along path~1 and path~2 indicated by the colored lines in (c).
(d3) Reconstructed $E$-$k$ map by integrating the two maps presented (d1) and (d2).
For the integration, the intensity ratio between the data from (d1) and (d2) was adjusted to be 1:0.8.
}
\label{fig3}
\end{center}
\end{figure}

\begin{figure}[t!]
\begin{center}
\includegraphics[width=1.0\columnwidth]{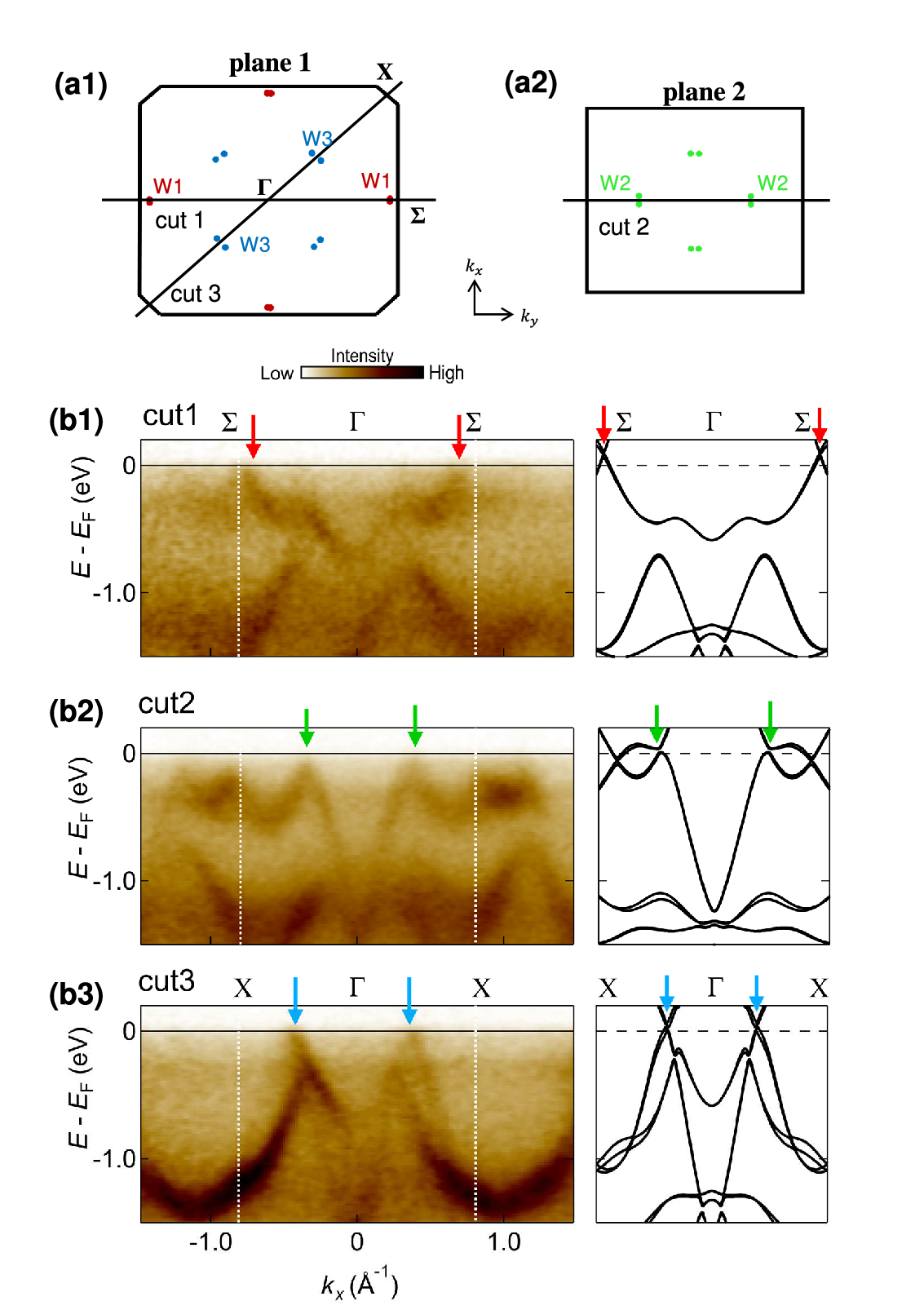}
\caption{
(a1), (a2) Distributions of Weyl nodes (W1, W2 and W3) in paramagnetic $R\rm{AlSi}$, predicted by DFT band calculations [see Fig.~\ref{fig1}(b)].
(b1)-(b3) SX-ARPES band maps cut along cut1-cut3 denoted by solid lines in (a) and taken from the reconstructed photoelectron intensity data presented in Fig.~\ref{fig2} and Fig.~\ref{fig3}.
See also Supplementary Note~3 for the dataset without the intensity reconstructions. 
}
\label{fig4}
\end{center}
\end{figure}

One might pose a question how the photoelectron intensity modulation arises in CeAlSi.
In fact, such an effect has been extensively studied in $\pi$ and $\sigma$ bands of graphite~\cite{Shirley_prb95,Daimon_jrsrp95}, referred to as $\it{zone}$-$\it{selection{\;} effect}$, which can be elucidated as a consequence of interference between photoelectron amplitudes from the orbitals of different sublattice atoms.
Thus, the zone-selection effect provides insight into the local atom environments as well as the orbital character forming the electronic bands~\cite{Matsui_jesrp14}.
We believe that a similar explanation can also be applied to CeAlSi, as its nonsymmorphic crystal structure contains sublattices.
It is noteworthy that a similar photoelectron modulation is also observed clearly in NdAlSi (see supplementary Note~2).
Therefore, the shared behavior in both materials likely indicates that the observed modulation of photoelectron intensity follows the zone-selection effect.
Hereafter we will refer to the modulation of photoelectron intensity as the zone-selection effect.

The intensity variation according to the zone-selection effect is also sensitive to momentum ($k$).
This is evident in the $E$-$k_x$ band maps on the $\Gamma_1$-plane and $\Gamma_2$-plane [Figs.~\ref{fig2}(b1) and \ref{fig2}(b2)].
For example, on the $\Gamma_2$-plane [Fig.~\ref{fig2}(b2)], bands with strong intensity are observed near the Fermi level ($E_{\rm{F}}$) and the $\Sigma$ points, but the intensity disappears around $\Gamma_2$ point.
Conversely, on the $\Gamma_1$-plane [Fig.~\ref{fig2}(b1)], the intensity of these bands diminishes around the $\Sigma$ points while are clearly observed near $\Gamma_1$ point.

The observed $k$ dependence of the photoelectron intensity distribution renders it rather difficult to determine the overall electronic band structures.
Nevertheless, even under such difficulty, our data covering multiple BZs enables us to compensate for the strong variations in the intensity caused by the zone-selection effect.
This is demonstrated in Fig.~\ref{fig2}(a3), where the FS maps on the $\Gamma_1$-plane [Fig.~\ref{fig2}(a1)] and $\Gamma_2$-plane [Fig.~\ref{fig2}(a2)] are integrated.
Here, one can now see the periodic pattern of the FSs including their intensity.
The same procedure to compensate for the intensity variations according to the zone-selection effect can also be applied to the band map [Fig.~\ref{fig2}(b3)], resulting in restored intensity and revealing the overall band structures.
The reconstructions of intensity distributions in the FS maps and band images demonstrate a good agreement with the electron structure obtained from DFT calculations [Figs.~\ref{fig2}(a4) and \ref{fig2}(b4)].

By utilizing the good $k_z$-resolution of SX photons~\cite{Strokov_jesrp03} in addition to the reconstruction of the intensity distributions overcoming the zone-selection effect, our data afford to visualize the overall band structures along high-symmetry $k$-paths including the $k_z$ axis [Fig.~\ref{fig3}(a)], which is crucial for experimentally tracing Weyl-cone dispersions interspersed in $k$-space.
Our data obtained by scanning $h\nu{'}$s contain multi-dimensional information, including the photoelectron intensity as a function of $E$, $k_x$ and $k_z$ [Fig. \ref{fig3}(b)].
Figures~\ref{fig3}(d1) and \ref{fig3}(d2) represent the band maps taken from the three-dimensional data along two equivalent $k$-paths [marked by the red and blue lines in Figs.~\ref{fig3}(b) and \ref{fig3}(c)].
Despite clear differences in intensity distributions along the two equivalent paths, the reconstruction procedure of the intensity allows us to compensate its variations and visualize the overall band dispersion including $k_z$-axis as clearly demonstrated in Fig.~\ref{fig3}(d3).

Finally, we apply this intensity analysis to trace the Weyl-cone dispersion (W1, W2 and W3) scattered across the $k$-space~\cite{gaudet2021weyl}.
To achieve this, we select the three $k$-paths on the two $k_x$-$k_y$ planes [plane~1 and plane~2 in Fig.~\ref{fig4}(a1) and (a2), see also Fig.~\ref{fig1}(b)] taken from the three-dimensional data in which intensity variation according to the zone-selection effect is reconstructed [Figs.~\ref{fig2}(a3) and \ref{fig3}].
Our experimental results along Cut1 [$\Sigma$-$\Gamma$-$\Sigma$ line, Fig.~\ref{fig4}(b1)] and Cut3 [$\rm{X}$-$\Gamma$-$\rm{X}$ line, Fig.~\ref{fig4}(b3)] validate the band dispersion of W1 and W3 as predicted by DFT calculation [the red and blue arrows in Figs~\ref{fig4}(b1) and (b3)].
W2 appears on the $k_x$-$k_y$ plane at $k_z$=0.3~$\rm{\AA}^{-1}$ according to the theory [green arrows in Fig~\ref{fig4}(b2)], which is experimentally observed in Fig.~\ref{fig4}(a3).

It is worth noting that the photoelectron intensity of the Weyl-cone dispersion is also sensitive to the BZ according to the zone-selection effect (see Supplementary~Note~3).
Therefore, ARPES measurements conducted solely within a single BZ would be insufficient to fully determine the electronic structures.
Especially in nonsymmorphic systems, where a similar zone-selection effect is expected, caution is required in experiments to fully determine the overall electronic structures.

The photoelectron intensity analysis demonstrated in this study is believed to be useful in such cases and proposes a new utility for tunable SX photons from synchrotron radiation.
The overall band structures show a good agreement with the DFT calculation, suggesting that these Weyl nodes is well located in the vicinity of $E_{\rm{F}}$.
The novel electronic states, therefore, should give a significant impact on macroscopic transport properties in the both paramagnetic and magnetic phases in CeAlSi and NdAlSi.

In conclusion, we observed the zone-selection effect on photoelectron intensity in the nonsymmorphic systems CeAlSi and NdAlSi, which makes it difficult to determine the overall band structure.
We overcome this difficulty covering multiple BZs thanks to synchrotron soft X-ray beams.
Our data reconstruct the photoelectron distributions, which unravels comprehensive information about the electronic band structures reproduced by band calculations including the Weyl-cone dispersion in the vicinity of $E_{\rm{F}}$.
Although our study is conducted in the paramagnetic phase, our results are important also for understanding the properties in magnetic phase.

We thank J.~Nag for supporting our SX-ARPES experiments.
This work is supported by JSPS KAKENHI Grant Number 24H01670, JP24K06943, 	23K23211, 23K17671, 	22H04483, JP22H01943, JP21H04652).
The synchrotron radiation experiments were performed with the approval of Japan Synchrotron Radiation Research Institute (JASRI)
(Proposals No.~2022A1434, 2022B1357, 2022A2060 and 2022B2106).

\bibliography{Morita_NdAlSi_sb1}

\begin{thebibliography}{10}

\bibitem{Weng_jpcm2016}
H. Weng, X. Dai, and Z. Fang, Journal of Physics: Condensed Matter {\bf 28},
  303001  (2016).

\bibitem{binghai_arcmp2017}
B. Yan and C. Felser, Annual Review of Condensed Matter Physics {\bf 8},  337
  (2017).

\bibitem{armitage_rmp2018}
N.~P. Armitage, E.~J. Mele, and A. Vishwanath, Rev. Mod. Phys. {\bf 90},
  015001  (2018).

\bibitem{Ding_rmp2021}
B.~Q. Lv, T. Qian, and H. Ding, Rev. Mod. Phys. {\bf 93},  025002  (2021).

\bibitem{Borisenko_prl14}
S. Borisenko, Q. Gibson, D. Evtushinsky, V. Zabolotnyy, B. B\"uchner, and R.~J.
  Cava, Phys. Rev. Lett. {\bf 113},  027603  (2014).

\bibitem{Xu_Science15_Dirac}
S.-Y. Xu, C. Liu, S.~K. Kushwaha, R. Sankar, J.~W. Krizan, I. Belopolski, M.
  Neupane, G. Bian, N. Alidoust, T.-R. Chang, H.-T. Jeng, C.-Y. Huang, W.-F.
  Tsai, H. Lin, P.~P. Shibayev, F.-C. Chou, R.~J. Cava, and M.~Z. Hasan,
  Science {\bf 347},  294  (2015).

\bibitem{Liu_NatureMat14}
Z.~K. Liu, J. Jiang, B. Zhou, Z.~J. Wang, Y. Zhang, H.~M. Weng, D. Prabhakaran,
  S.-K. Mo, H. Peng, P. Dudin, T. Kim, M. Hoesch, Z. Fang, X. Dai, Z.~X. Shen,
  D.~L. Feng, Z. Hussain, and Y.~L. Chen, Nature Materials {\bf 13},  677
  (2014).

\bibitem{Xu_Science15}
S.-Y. Xu, I. Belopolski, N. Alidoust, M. Neupane, G. Bian, C. Zhang, R. Sankar,
  G. Chang, Z. Yuan, C.-C. Lee, S.-M. Huang, H. Zheng, J. Ma, D.~S. Sanchez, B.
  Wang, A. Bansil, F. Chou, P.~P. Shibayev, H. Lin, S. Jia, and M.~Z. Hasan,
  Science {\bf 349},  613  (2015).

\bibitem{Lv_prx15}
B.~Q. Lv, H.~M. Weng, B.~B. Fu, X.~P. Wang, H. Miao, J. Ma, P. Richard, X.~C.
  Huang, L.~X. Zhao, G.~F. Chen, Z. Fang, X. Dai, T. Qian, and H. Ding, Phys.
  Rev. X {\bf 5},  031013  (2015).

\bibitem{Damascelli_rmp03}
A. Damascelli, Z. Hussain, and Z.-X. Shen, Rev. Mod. Phys. {\bf 75},  473
  (2003).

\bibitem{Sobota_rmp2021}
J.~A. Sobota, Y. He, and Z.-X. Shen, Rev. Mod. Phys. {\bf 93},  025006  (2021).

\bibitem{Zhang_NatureRev22}
H. Zhang, T. Pincelli, C. Jozwiak, T. Kondo, R. Ernstorfer, T. Sato, and S.
  Zhou, Nature Reviews Methods Primers {\bf 2},  54  (2022).

\bibitem{Nakatsuji_ANN2022}
S. Nakatsuji and R. Arita, Annual Review of Condensed Matter Physics {\bf 13},
  119  (2022).

\bibitem{Kuroda_NM2017}
K. Kuroda, T. Tomita, M.-T. Suzuki, C. Bareille, A.~A. Nugroho, P. Goswami, M.
  Ochi, M. Ikhlas, M. Nakayama, S. Akebi, R. Noguchi, R. Ishii, N. Inami, K.
  Ono, H. Kumigashira, A. Varykhalov, T. Muro, T. Koretsune, R. Arita, S. Shin,
  T. Kondo, and S. Nakatsuji, Nature Materials {\bf 16},  1090  (2017).

\bibitem{Liu_SciAdv19}
D.~F. Liu, A.~J. Liang, E.~K. Liu, Q.~N. Xu, Y.~W. Li, C. Chen, D. Pei, W.~J.
  Shi, S.~K. Mo, P. Dudin, T. Kim, C. Cacho, G. Li, Y. Sun, L.~X. Yang, Z.~K.
  Liu, S.~S.~P. Parkin, C. Felser, and Y.~L. Chen, Science {\bf 365},  1282
  (2019).

\bibitem{Hodovanets_prb18}
H. Hodovanets, C.~J. Eckberg, P.~Y. Zavalij, H. Kim, W.-C. Lin, M. Zic, D.~J.
  Campbell, J.~S. Higgins, and J. Paglione, Phys. Rev. B {\bf 98},  245132
  (2018).

\bibitem{Su_prb21}
H. Su, X. Shi, J. Yuan, Y. Wan, E. Cheng, C. Xi, L. Pi, X. Wang, Z. Zou, N. Yu,
  W. Zhao, S. Li, and Y. Guo, Phys. Rev. B {\bf 103},  165128  (2021).

\bibitem{yang2021noncollinear}
H.-Y. Yang, B. Singh, J. Gaudet, B. Lu, C.-Y. Huang, W.-C. Chiu, S.-M. Huang,
  B. Wang, F. Bahrami, B. Xu, {\it et~al.}, Physical Review B {\bf 103},
  115143  (2021).

\bibitem{sciadv.1603266}
S.-Y. Xu, N. Alidoust, G. Chang, H. Lu, B. Singh, I. Belopolski, D.~S. Sanchez,
  X. Zhang, G. Bian, H. Zheng, M.-A. Husanu, Y. Bian, S.-M. Huang, C.-H. Hsu,
  T.-R. Chang, H.-T. Jeng, A. Bansil, T. Neupert, V.~N. Strocov, H. Lin, S.
  Jia, and M.~Z. Hasan, Science Advances {\bf 3},  e1603266  (2017).

\bibitem{Sanchez:2020aa}
D.~S. Sanchez, G. Chang, I. Belopolski, H. Lu, J.-X. Yin, N. Alidoust, X. Xu,
  T.~A. Cochran, X. Zhang, Y. Bian, S.~S. Zhang, Y.-Y. Liu, J. Ma, G. Bian, H.
  Lin, S.-Y. Xu, S. Jia, and M.~Z. Hasan, Nature Communications {\bf 11},  3356
   (2020).

\bibitem{PhysRevB.107.035158}
R. Lou, A. Fedorov, L. Zhao, A. Yaresko, B. B\"uchner, and S. Borisenko, Phys.
  Rev. B {\bf 107},  035158  (2023).

\bibitem{PhysRevMaterials.7.L051202}
A.~P. Sakhya, C.-Y. Huang, G. Dhakal, X.-J. Gao, S. Regmi, B. Wang, W. Wen,
  R.-H. He, X. Yao, R. Smith, M. Sprague, S. Gao, B. Singh, H. Lin, S.-Y. Xu,
  F. Tafti, A. Bansil, and M. Neupane, Phys. Rev. Mater. {\bf 7},  L051202
  (2023).

\bibitem{Zhang:2023aa}
Y. Zhang, Y. Gao, X.-J. Gao, S. Lei, Z. Ni, J.~S. Oh, J. Huang, Z. Yue, M.
  Zonno, S. Gorovikov, M. Hashimoto, D. Lu, J.~D. Denlinger, R.~J. Birgeneau,
  J. Kono, L. Wu, K.~T. Law, E. Morosan, and M. Yi, Communications Physics {\bf
  6},  134  (2023).

\bibitem{Li:2023aa}
C. Li, J. Zhang, Y. Wang, H. Liu, Q. Guo, E. Rienks, W. Chen, F. Bertran, H.
  Yang, D. Phuyal, H. Fedderwitz, B. Thiagarajan, M. Dendzik, M.~H. Berntsen,
  Y. Shi, T. Xiang, and O. Tjernberg, Nature Communications {\bf 14},  7185
  (2023).

\bibitem{PhysRevB.109.035120}
A. Laha, A.~K. Kundu, N. Aryal, E.~S. Bozin, J. Yao, S. Paone, A.
  Rajapitamahuni, E. Vescovo, T. Valla, M. Abeykoon, R. Jing, W. Yin, A.~N.
  Pasupathy, M. Liu, and Q. Li, Phys. Rev. B {\bf 109},  035120  (2024).

\bibitem{Cheng:2024aa}
E. Cheng, L. Yan, X. Shi, R. Lou, A. Fedorov, M. Behnami, J. Yuan, P. Yang, B.
  Wang, J.-G. Cheng, Y. Xu, Y. Xu, W. Xia, N. Pavlovskii, D.~C. Peets, W. Zhao,
  Y. Wan, U. Burkhardt, Y. Guo, S. Li, C. Felser, W. Yang, and B. B{\"u}chner,
  Nature Communications {\bf 15},  1467  (2024).

\bibitem{suzuki2019singular}
T. Suzuki, L. Savary, J.-P. Liu, J.~W. Lynn, L. Balents, and J.~G. Checkelsky,
  Science {\bf 365},  377  (2019).

\bibitem{gaudet2021weyl}
J. Gaudet, H.-Y. Yang, S. Baidya, B. Lu, G. Xu, Y. Zhao, J.~A.
  Rodriguez-Rivera, C.~M. Hoffmann, D.~E. Graf, D.~H. Torchinsky, {\it et~al.},
  Nature materials {\bf 20},  1650  (2021).

\bibitem{sun2021mapping}
Y. Sun, C. Lee, H.-Y. Yang, D.~H. Torchinsky, F. Tafti, and J. Orenstein,
  Physical Review B {\bf 104},  235119  (2021).

\bibitem{Guoqing_prb18}
G. Chang, B. Singh, S.-Y. Xu, G. Bian, S.-M. Huang, C.-H. Hsu, I. Belopolski,
  N. Alidoust, D.~S. Sanchez, H. Zheng, H. Lu, X. Zhang, Y. Bian, T.-R. Chang,
  H.-T. Jeng, A. Bansil, H. Hsu, S. Jia, T. Neupert, H. Lin, and M.~Z. Hasan,
  Phys. Rev. B {\bf 97},  041104  (2018).

\bibitem{Muro_jsr2021}
T. Muro, Y. Senba, H. Ohashi, T. Ohkochi, T. Matsushita, T. Kinoshita, and S.
  Shin, Journal of Synchrotron Radiation {\bf 28},  1631  (2021).

\bibitem{Blaha2020}
P. Blaha, K. Schwarz, F. Tran, R. Laskowski, G.~K.~H. Madsen, and L.~D. Marks,
  J. Chem. Phys. {\bf 152},  074101  (2020).

\bibitem{Wu2017}
Q. Wu, S. Zhang, H.-F. Song, M. Troyer, and A.~A. Soluyanov, Computer Physics
  Communications {\bf 224},  405  (2018).

\bibitem{Shirley_prb95}
E.~L. Shirley, L.~J. Terminello, A. Santoni, and F.~J. Himpsel, Phys. Rev. B
  {\bf 51},  13614  (1995).

\bibitem{Daimon_jrsrp95}
H. Daimon, S. Imada, H. Nishimoto, and S. Suga, Journal of Electron
  Spectroscopy and Related Phenomena {\bf 76},  487  (1995), proceedings of the
  Sixth International Conference on Electron Spectroscopy.

\bibitem{Matsui_jesrp14}
F. Matsui, T. Matsushita, and H. Daimon, Journal of Electron Spectroscopy and
  Related Phenomena {\bf 195},  347  (2014).

\bibitem{Strokov_jesrp03}
V. Strocov, Journal of Electron Spectroscopy and Related Phenomena {\bf 130},
  65  (2003).

\end{thebibliography}
\bibliographystyle{prsty}

\title{\bf Supplementary Information:\\ 
Zone-selection effect of photoelectron intensity distributions \\in a nonsymmorphic system $R$AlSi ($R$: Ce and Nd)}

\author{Yusei Morita$^1$}
\author{K. Nakanishi$^2$} 
\author{T. Iwata$^{2,3}$} 
\author{K. Ohwada$^2$} 
\author{Y. Nishioka$^2$}
\author{T. Kousa$^2$}
\author{M. Nurmamat$^2$} 
\author{K. Yamagami$^4$} 
\author{A. Kimura$^{2,3,5}$} 
\author{T. Yamada$^6$} 
\author{H. Tanida$^6$} 
\author{Kenta Kuroda$^{2,3,5}$}

\affiliation{$^1$Department of Physics, Faculty of Science, Hiroshima University, Higashi-hiroshima, Hiroshima 739-8526, Japan}
\affiliation{$^2$Graduate School of Advanced Science and Engineering, Hiroshima University, Higashi-hiroshima, Hiroshima 739-8526, Japan}
\affiliation{$^3$International Institute for Sustainability with Knotted Chiral Meta Matter (WPI-SKCM$^2$), Hiroshima University, Higashi-hiroshima, Hiroshima 739-8526, Japan}
\affiliation{$^4$Synchrotron Radiation Research Center, Japan Atomic Energy Agency, Hyogo 679-5148, Japan}
\affiliation{$^5$Research Institute for Semiconductor Engineering, 1-4-2 Kagamiyama, Higashi-Hiroshima, Hiroshima 739-8527, Japan}
\affiliation{$^6$Liberal Arts and Sciences, Faculty of Engineering, Toyama Prefectural University, Izumi, Toyama 939-0398, Japan}

\maketitle

\newpage
\onecolumngrid
\subsection{The signature for $k_z$ dispersion in NdAlSi}
\begin{suppfigure}[H]
\begin{center}
\includegraphics[width=0.7\columnwidth]{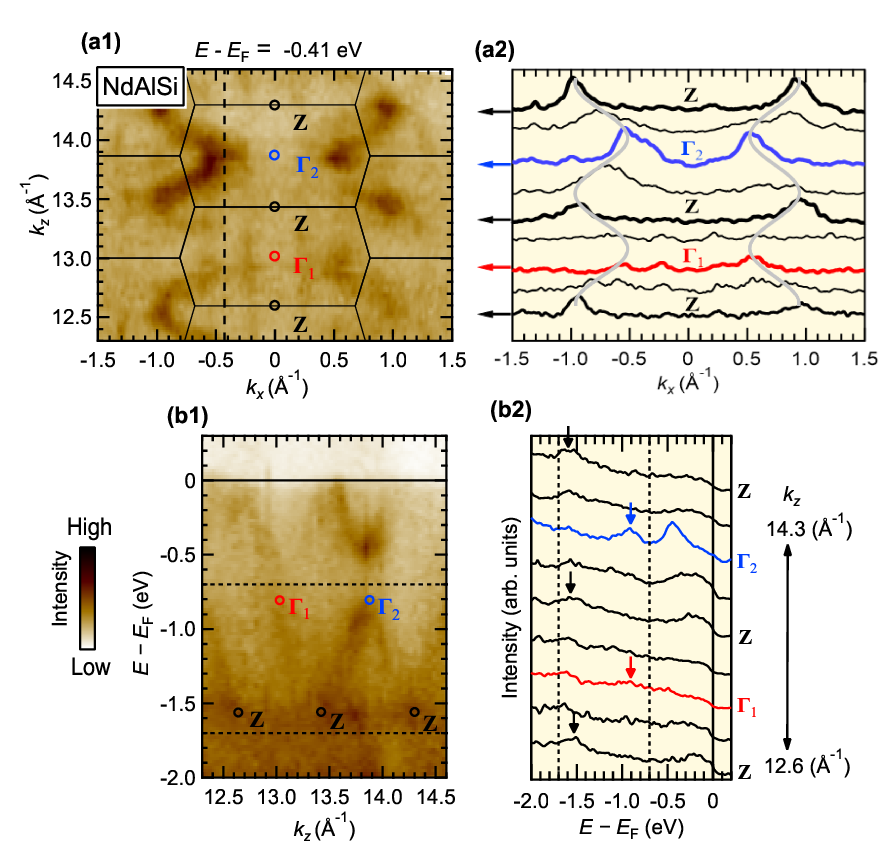}
\caption{
(a1) $k_z$-$k_x$ constant energy contour mapping of NdAlSi at $E-E_{\rm{F}}$=$-$0.41~eV, scanned by varying $h\nu{'}$s from 560 to 780~eV with 5~eV step.
The inner potential $V_0$=17~eV is used to convert the $h\nu$ dependency of the experimental bands to the corresponding $k_z$ dispersion.
The boundary of the BZs are drawn by solid lines.
(a2) The momentum distribution curves (MDCs) of (a1) at representative $k_z$-levels.
The MDCs cutting the high-symmetry $k_z$ point are highlighted by the bold lines.  
(b1) The band map cut on the $E$-$k_z$ plane at $k_{x}$=$-$0.43$~\rm{\AA^{-1}}$ [the dashed line in (a1)].
(b2) The representative energy distribution curves.
The periodic $k_z$-dispersion is traced by the arrows.  
}
\label{figS1}
\end{center}
\end{suppfigure}

Supplementary~Figure~\ref{figS1}(a1) presents the $k_{z}$-$k_{x}$ constant energy contours of NdAlSi at $k_y$=0~${\rm{\AA}}^{-1}$ at $E-E_{\rm{F}}$=$-$0.41~eV recorded with varying $h\nu{'}$s from 560 to 780~eV.
This data covers multiple BZs including a number of BZs in NdAlSi as observed in CeAlSi [Fig.~1 in the main text].
We observe a chain like $k_z$ dispersion around $\Sigma$ points, which unambiguously indicates that the observed band features belong to bulk bands.
The signature of this $k_z$ dispersion can be observed more distinctly in Supplementary~Fig.~\ref{figS1}(a2) where we show representative momentum distribution curves (MDCs).
Supplementary~Figures~\ref{figS1}(b1) and (b2), respectively, display the $E$-$k_z$ image and its energy momentum cuts (EDCs). 
The peak position in the EDCs varies with $k_z$, following a 2${\pi}/c$ periodicity (gray lines).
This exhibits a band folding with a periodicity of ${\Delta}k_z{\sim}$0.86~${\rm{\AA}}^{-1}$ that is consistent with 2${\pi}/c$.
\subsection{The zone-selection effect in NdAlSi}
\begin{suppfigure}[h!]
\begin{center}
\includegraphics[width=0.9\textwidth]{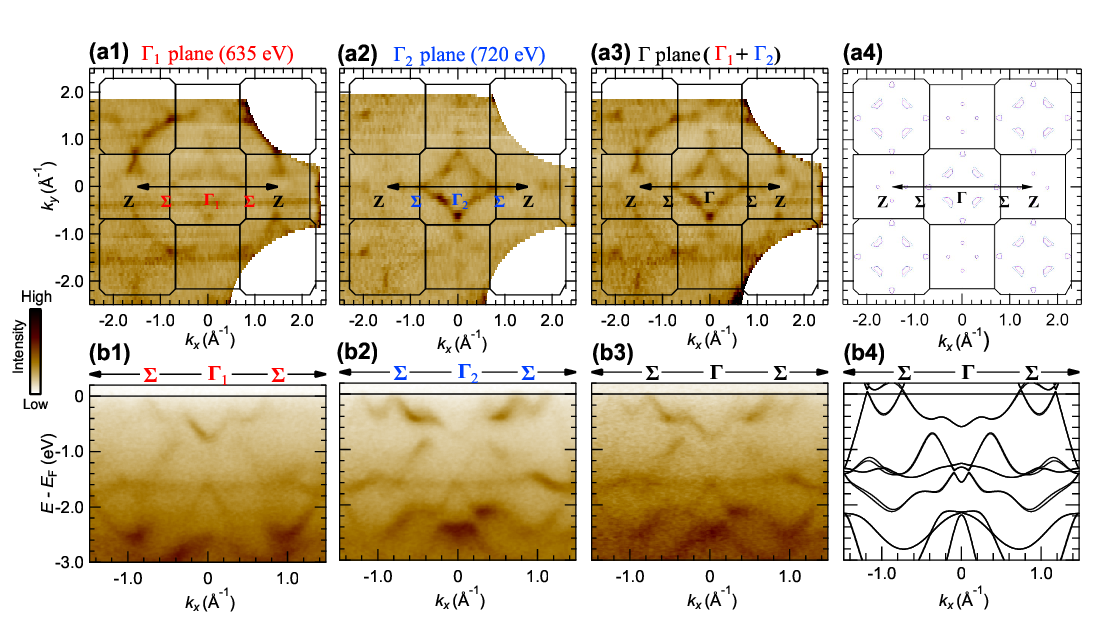}
\caption{
(a1), (a2) The results of the Fermi surface mapping in NdAlSi on the $k_x$-$k_y$ planes cutting $\Gamma_1$ ($h\nu$=635~eV) and $\Gamma_2$ ($h\nu$=720~eV).  
(a3) The reconstructed photoelectron intensity distribution obtained by summing the intensities of the FS maps [(a1) and (a2)], where the data from (a1) and (a2) were weighted with ratios of 1:0.8.
(a4) The calculated Fermi surface at $\Gamma$.
(b1)-(b3) The photoelectron distributions cut along $\Sigma$-$\Gamma$-$\Sigma$ line for data shown in (a1)-(a3), compared to the calculated band dispersions shown in (b4).
We shift the overall DFT bands in energy of 80 meV to obtain the best fit.
}
\label{figS2}
\end{center}
\end{suppfigure}%

We observe the zone-selection effect not only in CeAlSi but also in NdAlSi.
This can be already recognized in the MDCs as shown in Supplementary Fig.~\ref{figS1}(a2).
The shape of the MDC at the two $\Gamma$ points ($\Gamma_1$ and $\Gamma_2$) differ considerably, despite these two $k_z$-levels being equivalent in BZ, while the MDCs at each $\rm{Z}$ point exhibit similar shapes.

Supplementary~Figures~\ref{figS2}(a1-a2) and (b1-b2) display the $k_x$-$k_y$ Fermi surface contours and $E$-$k_x$ band maps in NdAlSi.
Like the zone-selection rule observed in CeAlSi [Fig.~2 in the main text], we observe the different intensity distributions on the $\Gamma_1$-plane [Supplementary Figs.~\ref{figS2}(a1) and (b1)] and $\Gamma_2$-plane [Supplementary~Figs.~\ref{figS2}(a2) and (b2)].
Although the zone-selection effect dominates the photoelectron intensity distribution in the energy dispersions, it can be compensated in Supplementary~Fig.~\ref{figS2}(a3) and \ref{figS2}(b3) where the FS maps on the $\Gamma_1$-plane and $\Gamma_2$-plane are integrated.

\subsection{The zone-selection effect in Weyl-cone dispersion}
\begin{suppfigure}[h!]
\begin{center}
\includegraphics[width=0.9\textwidth]{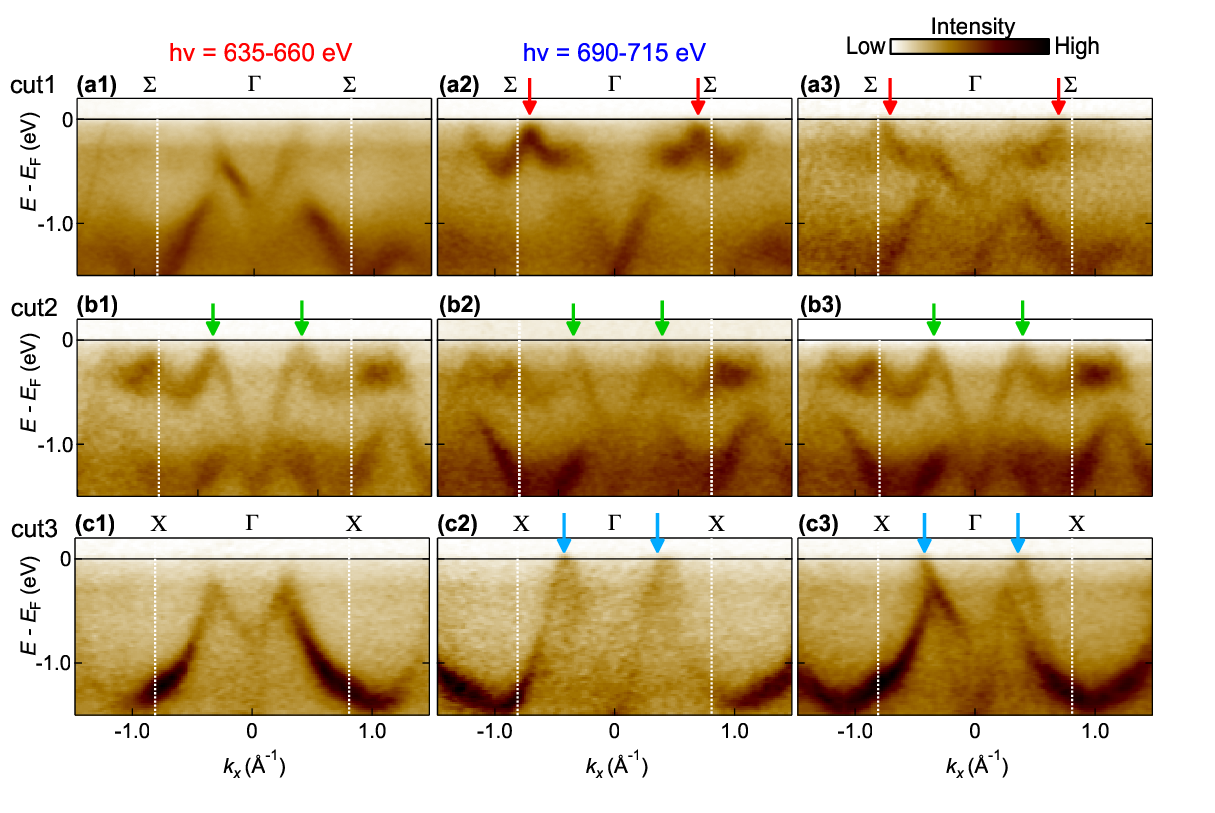}
\caption{
(a1)-(c1) and (a2)-(c2) SX-ARPES band maps along the equivalent momentum paths acquired with different photon energies (see Fig.~3 in the main text).
(a3)-(c3) The reconstructed photoelectron intensity maps.
The Weyl-cone dispersion are denoted by arrows. 
}
\label{figS3}
\end{center}
\end{suppfigure}
In Fig. 4 of the main text, we present the dispersion of the Weyl-cone in CeAlSi investigated by SX-ARPES. All of these data are obtained by summing up photoelectron intensities acquired in different zones to compensate for the zone-selection effect.
The raw data without compensation are shown in Supplementary Figs. \ref{figS3}(a1)-(c1) and \ref{figS3}(a2)-(c2).
We used the photon energy range of 635-660 eV for Supplementary Figs. \ref{figS3}(a1)-(c1) and 690-715 eV for Supplementary Figs. \ref{figS3}(a2)-(c2).
A significant zone-selection effect can be observed, particularly in cut1 and cut3.
The photoelectron intensity of the Weyl-cone dispersion is almost absent in Supplementary Figs. \ref{figS3}(a1) and \ref{figS3}(c1), while it is enhanced in the other Brillouin zone in Supplementary Figs. \ref{figS3}(a2) and \ref{figS3}(c2). 
From this result, it can be inferred that caution is needed when determining the dispersion of electronic bands solely based on ARPES measurements conducted within a single Brillouin zone.

\end{document}